# Intracavity THz generation using a thin lithium niobate plate in a compact Kerr-lens mode-locked Yb:CALGO bulk oscillator

**Mohsen Khalili,[1,*] Jokūbas Pimpė,[2] Yicheng Wang,[1] Julius Vengelis,[2] Kore Hasse,[3] Sergiy Suntsov,[3] Detlef Kip,[3] and Clara J. Saraceno[1]**

[1]*Photonics and Ultrafast Laser Science (PULS), Ruhr-Universität Bochum, Universitätsstraße 150, 44801 Bochum, Germany*

[2]*Vilnius University Laser Research Center, Saulėtekio Ave. 10, Vilnius, LT-10223, Lithuania*

[3]*Experimental Physics and Material Sciences, Helmut-Schmidt-Universität, Holstenhofweg 85, 22043 Hamburg, Germany*

**Mohsen.Khalilikelaki@ruhr-uni-bochum.de*

**Abstract:** We demonstrate intracavity terahertz (THz) generation via optical rectification in a 50-µm-thick lithium niobate crystal placed inside a compact diode-pumped Kerr-lens mode-locked (KLM) Yb:CALGO bulk oscillator. The oscillator operates at a repetition rate of 85 MHz and delivers 83-fs pulses with up to 71 W of average intracavity power, obtained with only 21.4 W of low-cost multimode diode pump power. We generate single-cycle THz pulses with a spectrum extending up to 3 THz, detected by electro-optic sampling with 60 dB dynamic range within 156 s of measurement time (313 averaged traces) and up to 120 µW of THz average power. This work combines the high damage threshold, power-handling capability, and cost-effectiveness of thin LN plates with simplicity, compactness, and low-cost multimode diode-pumped solid-state bulk lasers, offering an attractive alternative for high-repetition-rate THz time-domain spectroscopy systems.

## 1. Introduction

Broadband THz sources driven by ultrafast lasers are the cornerstone of THz time-domain spectroscopy (THz-TDS), which has become an indispensable tool in fields ranging from fundamental solid-state physics and molecular spectroscopy to non-destructive testing and imaging [1, 2]. Many imaging and spectroscopy applications based on THz time-domain spectroscopy (TDS) benefit from higher THz average power at MHz repetition rates, which enables fast data acquisition, low-noise detection schemes, and high dynamic range (DR).

In this context, the rapid progress of high-average-power ultrafast ytterbium-based laser technology has enabled remarkable advances in high-power THz generation over the last decades. Most of this progress has been realized using optical rectification (OR) in the regime of tens to hundreds of kHz repetition rates where mJ level pulse energies are nowadays becoming widely commercially available, with THz power levels now approaching the watt-level from both plasma- and OR-based sources [3, 4]. For example, in [4] 643 mW of average power has been demonstrated at 40 kHz using OR in the tilted-pulse-front (TPF) method in lithium niobate (LN) and OR in organic crystals has demonstrated >10 mW of average power at 100 kHz [5]. At higher repetition rates, conversion efficiencies reported are significantly lower: for instance a milliwatt-class broadband THz source has been demonstrated by driving OR in gallium phosphide (GaP) with a 112-W thin-disk laser oscillator [6], and up to 66 mW of THz average power was obtained by OR in LN using the TPF approach [7] with the same class of driving laser [8]. The difficulty at high repetition rates originates from the scaling of OR efficiency with peak intensity: increasing the repetition rate into the multi-MHz regime at constant (high) average power results in a drop in efficiency. Maintaining efficient conversion, on the other hand, requires driving average powers that are technically unavailable or strongly prohibitive; or generate excessive thermal load or damage. We note that photoconductive THz emitters are known to operate efficiently at low driving pulse energies and have recently reached milliwatt-level average powers at tens of MHz [9, 10]. Although this work is very promising, standard photoconductive emitters for this repetition-rate region still require specialized fabrication that constrains performance and keeps them costly, and versions compatible with high average powers are not yet commercially available. Other strategies, such as THz-emitting nonlinear metasurfaces, are also currently being explored [11, 12], but their capability for high-power THz generation remains far below that of bulk crystals so far.

An elegant alternative to exploit the wide availability and flexible parameter ranges of nonlinear-crystal emitters while circumventing the difficulties of operating at very high repetition rate is to exploit resonant enhancement, placing the nonlinear crystal inside a cavity where the driving light builds up over many roundtrips (RT) so that the circulating average power is significantly higher than the coupled-out power. This resonant-enhancement strategy has been used successfully for the generation of XUV light via intracavity high-harmonic generation, both in passive enhancement cavities [13] and directly inside mode-locked thin-disk oscillators [14]. Early demonstrations of intracavity THz generation used photocurrent transients in semiconductor structures inside Ti:sapphire [15] and fiber [16] oscillators, actively controlled passive enhancement cavities [17], and OR in zinc telluride inside a Ti:sapphire oscillator [18], but remained restricted to microwatt-level THz powers. With regards to Yb-lasers, OR in a GaP crystal placed inside a Kerr-lens mode-locked (KLM) diode-pumped Yb:CALGO bulk oscillator yielded up to 150 μW of THz power with spectra extending to 5.5 THz, using only 7 W of diode pump power [19]. However, the power scalability of GaP-based intracavity sources is known to be severely constrained by multi-photon absorption and the resulting free-carrier absorption and resulting thermal effects: at high average power, this induces strong thermal lensing that acts both as a diffraction loss mechanism for the cavity as well as a power modulator that can potentially destabilizes mode-locking [20], so operation at significantly higher intracavity power cannot be expected with this material. Furthermore, GaP is overall an expensive crystal far from mass production.

LN is in many aspects a better candidate to overcome these limitations compared with GaP: it offers a large effective nonlinear coefficient for THz generation and a wide bandgap, which strongly suppresses multi-photon absorption at 1-μm pump wavelengths and results in a high damage threshold [21]. Moreover, its power-handling capability has been proven at driving average powers ranging from hundreds of watts up to nearly 2 kW [4, 22, 23]. The large group-velocity mismatch between the near-infrared pump and the generated THz wave in LN is commonly circumvented using the TPF technique [7], which is, however, alignment-sensitive and challenging for intracavity operation. Instead, the use of very thin LN (thin-LN) plates restores collinear velocity matching over the crystal length. Furthermore, LN wafers are mass-produced, widely accessible, and cost-effective (typically ~300 USD per wafer). Following this approach, we recently demonstrated 1.3 mW of THz average power from a 50-μm-thick LN plate placed inside a modelocked thin-disk laser operating at 264 W of intracavity power [22]. In a complementary approach, the same thin-LN concept was applied in a passive enhancement cavity, where OR was driven at a record 1.9 kW of average power at 93 MHz, generating milliwatt-level THz pulses [23]. These results establish thin plates of LN as excellent low-loss nonlinear material for resonantly enhanced THz generation. Nevertheless, both thin-disk oscillators and passive enhancement cavities remain complex and highly specialized expensive systems that require significant expertise and/or complex locking electronics to maintain passive enhancement, limiting their accessibility for widespread use. Moreover, both platforms are inherently sensitive to intracavity loss: the power buildup in a passive enhancement cavity collapses when even sub-percent additional loss is introduced; in [23], for example, the insertion of the AR-coated thin-LN plate alone reduced the cavity gain from 328 to 240.

Likewise, the low RT gain of thin-disk oscillators demands low-loss intracavity elements. A diode-pumped bulk oscillator, in contrast, typically provides higher RT gain and can therefore tolerate substantially higher intracavity losses, relaxing the requirements on the nonlinear crystal, its antireflection (AR) coatings, and any additional intracavity optics.

In this work, we extend the thin-LN intracavity approach to a simple, compact, and low-cost solid-state bulk oscillator pumped by a multimode diode. Yb:CALGO (Yb:$CaGdAlO_4$) is an excellent gain medium for this purpose, combining a broad and smooth gain spectrum with a low quantum defect and reasonable thermal conductivity. We place a 50-µm-thick LN plate inside a KLM Yb:CALGO oscillator operating at 85 MHz and demonstrate stable mode-locked operation with 83-fs pulses at up to 71 W of intracavity average power, using only 21.4 W of diode pump power. The footprint of the THz-TDS system is 30 cm × 60 cm, as shown on Fig. 1e. We generate single-cycle THz pulses with a bandwidth of ~3 THz and up to 120 µW of THz average power collected from one side of the crystal, with a near-quadratic power scaling without significant saturation that indicates substantial headroom for further improvement in future implementations at higher intracavity power.

## 2. Experimental setup

The experimental setup is shown in Fig. 1a. The laser oscillator is based on a 2.5-mm-long Yb(7.5 at.%):CALGO crystal with AR coatings on both sides, mounted in a water-cooled copper holder. The gain medium is pumped by a commercial multimode fiber-coupled diode laser at 975 nm delivering up to 27 W from a 105-µm-core fiber. The pump beam is imaged into the gain medium with two aspheric lenses (f = 50 mm and f = 75 mm), resulting in a pump spot diameter of 165 µm. In a separate low-power transmission measurement, ~97% of the incident pump power was absorbed in a single pass through the crystal. A pair of concave mirrors (radius of curvature, RoC = 300 mm) forms the laser mode in the gain medium. To accurately design the mode-locked cavity, we performed standard ABCD-matrix calculations coupled with an iterative Kerr-lens element, as described in [24]. Fig. 1b shows the calculated beam caustic as a function of intracavity peak power (logarithmic color scale). According to ABCD-matrix calculations, the laser mode diameter in the gain medium is ~160 µm in continuous-wave operation and ~300 µm under mode-locking. After mode-locking is initiated, the gain crystal is translated slightly out of the focus to maintain an appropriate Kerr-lens strength as the intracavity peak power is scaled. The pump spot was intentionally chosen smaller than the laser mode to ensure efficient pump absorption and promote stable single-transverse-mode operation.

Since the goal of this work is to scale the intracavity power rather than the extracted output power, the intracavity mode was deliberately designed to be relatively large to limit intracavity nonlinearities and maximize intracavity peak power. A total RT group delay dispersion of $-2000\ fs^2$ is introduced by highly dispersive (HD) mirrors to compensate for the positive dispersion and self-phase modulation contributions of the gain medium and the thin LN plate. Stable KLM is initiated and maintained with a 1.8-mm hard aperture (HA). Fig. 1c shows the corresponding beam radius at the position of the HA versus intracavity peak power, and Fig. 1d the resulting RT transmission through the 1.8-mm aperture. A 1% output coupler (OC) provides output beams for laser diagnostics and for the electro-optic sampling (EOS) probe; in future implementations one can consider minimizing the loss of this mirror for obtaining more intracavity enhancement.

The THz emitter is a 50-µm-thick, x-cut, 5 mol.% MgO-doped congruent LN plate (Fig. 1f), AR-coated on both sides for the laser wavelength, fabricated as described in [22]. Both the driving pulse and the generated THz field are polarized along the optic axis of the crystal in order to exploit the largest nonlinear coefficient $d_{33}$. The thin-LN plate is placed near the second intracavity focus formed by two concave mirrors (RoC = 200 mm and 100 mm) and is slightly offset from the beam waist so that the crystal position coincides with the focal plane of the first off-axis parabolic mirror ($OAP_1$, f = 101.6 mm) for optimal THz collection. Displacing the plate rather than the parabola is dictated by the space available in the resonator: with a shorter-focal-length parabola the collected THz power was too low to be measured, and the 101.6-mm focal length used here cannot be accommodated by moving $OAP_1$ further away from the crystal, while lengthening the telescope formed by the two concave mirrors would have required curved mirrors that were not available in our laboratory at the time and would also have increased the spot size on the crystal, possibly decreasing the conversion efficiency. $OAP_1$ is a commercially available gold-coated parabola with a through-hole at its center; it is placed directly inside the resonator, with the intracavity laser beam passing through the hole. The calculated beam diameter on the thin-LN is 300 µm ($1/e^2$ intensity). The other OAPs have a focal length of 50.8 mm.

For THz detection, a set of four off-axis parabolic mirrors collects and refocuses the generated THz beam at two focal positions. A mechanical chopper (1067 Hz) placed at the first focus modulates the THz beam for low-noise lock-in detection. At the second focus, the THz pulses are detected by conventional EOS in a 3-mm-thick GaP crystal using the 1% output-coupled beam as probe. Although such a thick crystal is known to limit the detection bandwidth, no thinner detection crystal was available at the time of the experiment. The delay between THz and probe pulses is scanned with a fast mechanical delay operating at 1 Hz over a 15-ps window, enabling the averaging of 313 traces in 156 s. The scan rate was chosen to ensure robust lock-in detection, as discussed in Sect. 4.

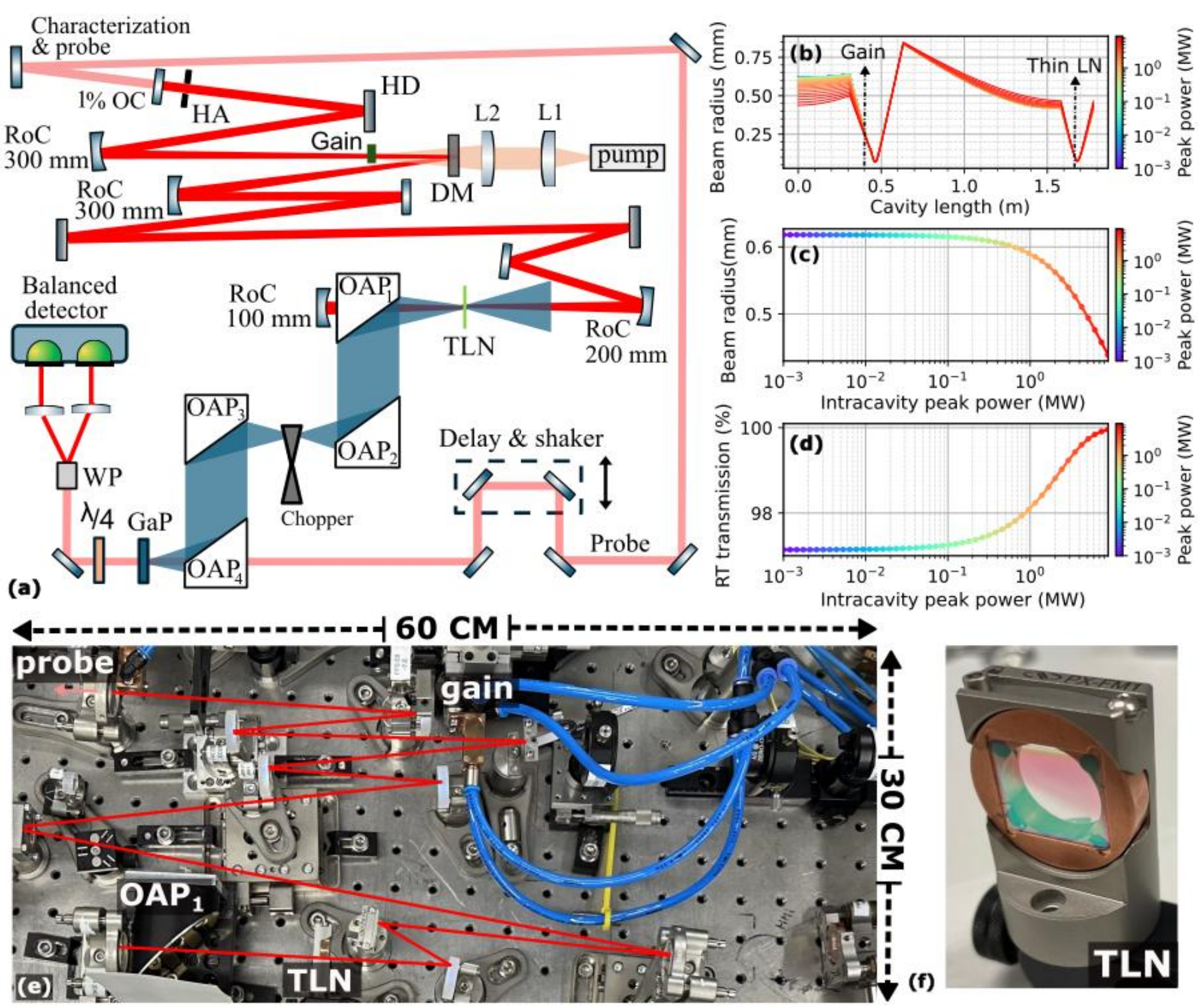

**Fig. 1** (a) Schematic of the experimental setup. DM, dichroic mirror; L1, L2, pump aspheric lenses; WP, Wollaston prism; λ/4, quarter-wave plate. The generated THz beam is collected by the intracavity holed mirror $OAP_1$, modulated by a mechanical chopper at an intermediate focus, and refocused on a 3-mm-thick GaP crystal for EOS with a balanced detection scheme. (b) Calculated beam caustic of the resonator, with the colormap indicating the intracavity peak power on a logarithmic scale; (c) beam radius at the position of the HA and (d) calculated RT transmission through the 1.8-mm HA, both as a function of intracavity peak power (same colormap as in (b)). (e) Photograph of the laser setup (footprint 30 cm × 60 cm) with the implemented thin-LN plate; (f) the thin-LN plate in a copper mount.

## 3. Results and discussion

### 3.1 Driving laser performance

The resonator is designed for operation at a repetition rate of 85 MHz. In the initial configuration, 57 W of intracavity average power is reached with 8.4 W of diode pump power without the thin-LN plate, and 9.3 W of diode pump power with the thin-LN plate inserted, demonstrating that the additional intracavity loss introduced by the 50-µm plate is small. The laser compensates for this additional loss with only about 10% more pump power, which illustrates the advantage of the high-gain bulk architecture. Fig. 2a,b compares the autocorrelation traces and optical spectra measured with and without the thin-LN inside the cavity, respectively. Assuming a sech$^2$ pulse shape, the pulse duration is 83 fs with the thin-LN and 82 fs without, while the optical spectrum, centered at ~1067 nm, has a full width at half maximum of 15.4 nm (15.5 nm without thin-LN), corresponding to a time-bandwidth product of ~0.34, close to the transform limit (0.315). The near-identical pulse parameters in both configurations confirm that the thin-LN plate does not significantly alter the soliton pulse formation nor introduces significant nonlinear loss, in clear contrast to intracavity GaP [20,22]. Mode-locking is based on the hard-aperture Kerr-lens mechanism and is initiated by slightly translating one of the cavity mirrors, typically the end mirror, after which it is maintained without further adjustment, and both the initiation and the stability of the mode-locked operation are unchanged by the insertion of the thin-LN plate. The oscillator runs in ambient air, without any enclosure, purging, or active stabilization. We expect that with simple engineering steps and in a sealed enclosure, the system can operate very reliably.

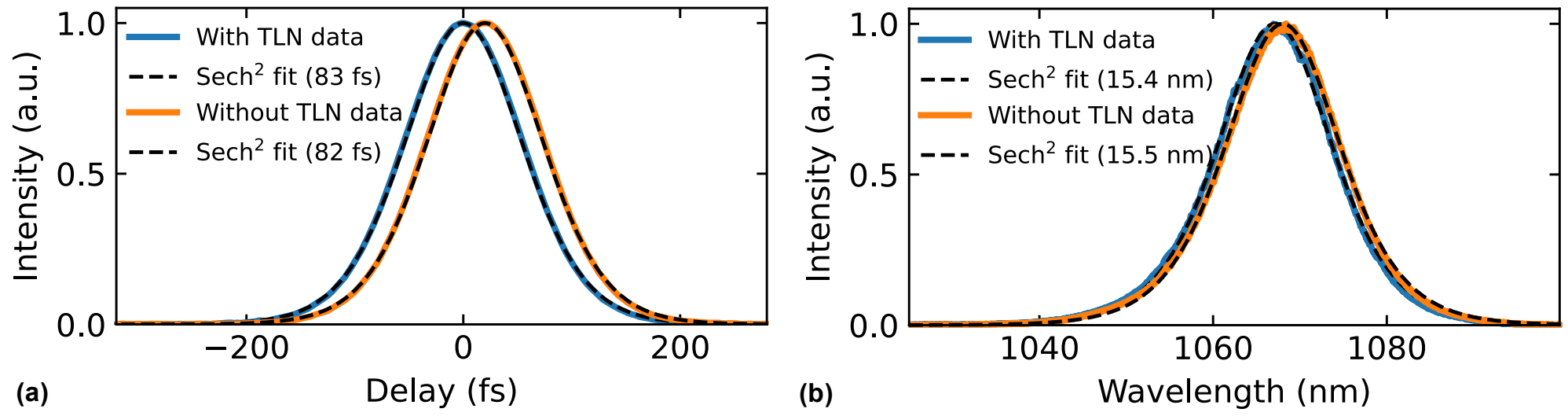

**Fig. 2** (a) Intensity autocorrelation traces of the mode-locked driving laser with and without the thin-LN plate inside the cavity, with sech² fits yielding 83 fs and 82 fs pulse duration, respectively (the trace without thin-LN is slightly offset horizontally for visibility). (b) Corresponding optical spectra with sech² fits of 15.4 nm and 15.5 nm bandwidth.

### 3.2 THz generation and detection

Fig. 3a shows the THz trace in time domain reconstructed by EOS, averaged over 313 traces (156 s acquisition time), together with a dark trace recorded with the THz beam blocked. A clean single-cycle THz transient is observed, followed by weak replicas of the pulse that originate from the internal reflection of the generated THz field inside the 50-µm-thick plate, which acts as a Fabry–Perot etalon at THz frequencies, and by the characteristic long-lived ringing caused by resonant absorption and re-emission of the THz field by water vapor [25] in the uncontrolled ambient atmosphere (~70% relative humidity, ~50 cm of THz propagation from generation to detection). The corresponding power spectrum, shown in Fig. 3b, extends up to 3 THz with a peak DR of 60 dB.

We simulate the THz generation and detection chain (dashed red line in Fig. 3b) and compare the result with the measured spectrum**Fig. 3**. The simulation follows the approach employed in our previous intracavity and cavity-enhanced THz studies [22,23]: the THz field generated by OR in the 50-µm thin-LN plate is calculated by solving the one-dimensional coupled wave equations in the frequency domain, in a similar way to [26], using the measured driving pulse parameters and literature values for the optical constants of LN in the THz range [27]. The propagation of the generated field to the detection crystal accounts for the frequency-dependent diffraction of the THz beam and the spectral filtering introduced by the through-hole and the limited collection aperture of the parabolic mirror [28]. Detection is modeled by the EOS response function of the 3-mm-thick GaP crystal [28], which includes the mismatch between the THz phase velocity and the group velocity of the probe pulse, the THz absorption of GaP, and the finite probe pulse duration. In addition, the dispersion of the probe pulse in the thick GaP crystal is also considered. Finally, the first four internal reflections of the THz pulse in the thin-LN plate are included as echoes in the simulated time-domain trace, with delays and amplitudes determined by the THz RT time in the plate and the Fresnel reflection coefficients of its interfaces; higher-order reflections are below the noise floor and out of our scan range and are neglected. Apart from an overall normalization of the spectral amplitude, the simulation contains no free fit parameters. For comparison, Fig. 3b also shows the simulated spectrum of the THz field as generated in the thin-LN plate (dashed green line), including the etalon echoes but none of the detection effects. It extends far beyond the detected spectrum, which shows that the measured bandwidth is limited by the detection chain rather than by the generation process itself. To separate the two contributions of the detection chain, Fig. 3b also shows the simulated spectrum after the THz collection optics alone (dash-dotted purple line), that is, after the through-hole and the finite collection aperture of $OAP_1$ but before the EOS response. Because the far-field divergence of the THz beam scales inversely with frequency, the finite collection aperture truncates mainly the strongly divergent low-frequency components, whereas the central through-hole, which has a diameter of 3 mm and subtends only about 0.85° as seen from the crystal, removes a small fraction of the on-axis power at all frequencies. Accordingly, the collection optics attenuate the spectrum predominantly at low frequencies, by about 17 dB at 0.5 THz, 6 dB at 2 THz and 3 dB at 3 THz, and by less than 1 dB above 5 THz. The pronounced roll-off of the detected spectrum above 2 THz, in contrast, is dominated by the EOS response of the 3-mm-thick GaP crystal, which attenuates the spectrum by about 5 dB at 2 THz, 21 dB at 3 THz and 34 dB at 5 THz. We note that for our simulation using the model presented in [26] we use optical constants measured on congruent LN, which extend only to approximately 2 THz [29]. Above this range, both the refractive index and the absorption coefficient are extrapolations of the fitted functions and contain no resonance features, but a smoothly rising absorption. For our crystal cut and polarization, the TO phonon peak that influences our spectrum is the one near 7.4 THz [30], but the absorption rises already at frequencies significantly below it, eventually limiting, even for very thin crystals, the generation of very broad THz spectra. The absorption coefficient used in the simulation is shown in the inset of Fig. 3a, where the range over which it is extrapolated is marked explicitly.

The simulation reproduces the measured spectrum well up to approximately 2 THz. In particular, it reproduces the quasi-periodic modulation of the spectrum, which originates from the spectral interference between the main THz transient and its etalon echoes: the modulation period is given by the inverse of the RT time of the THz pulse in the thin plate and is thus a direct signature of the 50-μm crystal thickness. Varying the LN thickness in the simulation shows that the measurement is consistent with thicknesses between 48 and 50 μm, for which the simulated spectra reproduce both the position of the dip near 2 THz, which marks the frequency at which the phase mismatch between the driving pulse and the generated THz field accumulates to $2\pi$ over the crystal length, and the echo-induced modulation equally well; a finer distinction is not possible here, since the 15-ps scan window limits the spectral resolution to approximately 67 GHz. Above ~2 THz, the measured spectrum rolls off faster than predicted by the simulation. We attribute this to a combination of effects that are only partially captured by the model. On the detection side, water-vapor absorption, which is not included in the simulation, exhibits increasingly dense and strong lines above 1.5 THz in the unpurged beam path [25]. On the generation side, precise data for the THz refractive index and absorption of LN beyond ~2 THz, and their dependence on the crystal temperature under intracavity operation, are scarce [27, 29].

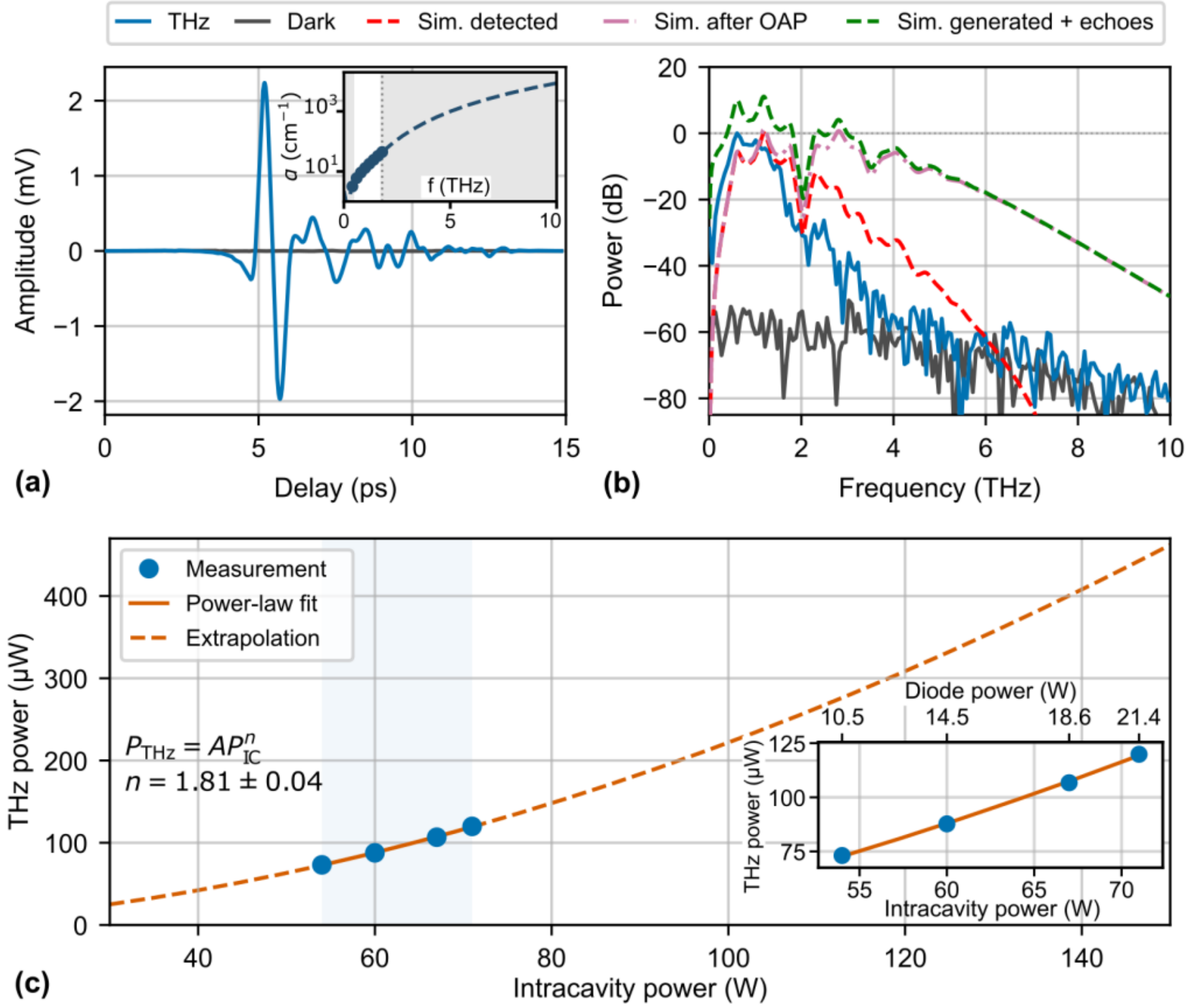

**Fig. 3** THz characterization: (a) Averaged THz time-domain trace (blue) and dark trace (gray) reconstructed by EOS, averaged over 313 scans in 156 s; the inset shows the THz intensity absorption coefficient of LN used in the simulation, with the range outside the measured interval from 0.4 to 1.8 THz shaded to indicate extrapolation; (b) corresponding power spectrum (blue), exhibiting a bandwidth of ~3 THz and a peak DR of 60 dB, together with the simulated spectrum generated in the thin-LN plate (dashed green), the simulated spectrum after the THz collection optics (dash-dotted purple), and the simulated detected spectrum including in addition the EOS response of the GaP crystal (dashed red); the three simulated curves share one common normalization, given by the peak of the simulated detected spectrum, while the measured and dark spectra are normalized to the peak of the measured spectrum, so that the simulated curves can exceed 0 dB; (c) THz average power (one side) versus intracavity power, with a power-law fit yielding n = 1.81 ± 0.04; the dashed line extrapolates the fit beyond the measured range (shaded region), and the inset shows a zoom into the measured data versus intracavity power (bottom axis) and diode pump power (top axis).

To investigate the power scaling behavior of the source, the diode pump power increased from 10.5 W to 21.4 W, raising the intracavity average power from 54 W to 71 W while maintaining stable single-pulse mode-locked operation. At the maximum intracavity power of 71 W, the peak fluence on the thin-LN is ~2.4 mJ/cm² and the peak intensity is ~25 GW/cm², well below the ~100 GW/cm² at which reliable intracavity operation of thin LN has already been demonstrated [22]. The THz average power is measured with a calibrated THz power meter (THz20, SLT Sensor- und Lasertechnik GmbH, calibrated by the German national metrology institute PTB). A 2-mm-thick germanium plate and a sheet of white paper block the residual pump and parasitic second-harmonic light, while the transparent protective cover attached to the sensor head shields the detector and reduces the influence of ambient air fluctuations. The measured THz power transmissions of the three elements are 20%, 76%, and 90%, respectively; all reported THz powers are corrected for the combined transmission of

~14%. The THz average power increases from 73 μW to 120 μW, as shown in Fig. 3c. A fit of the form $P_{THz} = A \cdot P_{IC}^{n}$ yields an exponent of n = 1.81 ± 0.04, close to the quadratic dependence expected for a second-order nonlinear process. The slight deviation from n = 2 may indicate the onset of weak saturation mechanisms. Crucially, however, we observed no roll-over, instability, or damage up to the maximum available pump power, which shows that the THz power is currently limited only by the available intracavity power and not by the nonlinear crystal. A moderate heating of the plate is nevertheless expected, which slightly increases the THz absorption of the material and contributes to the uncertainty of the simulated spectrum at higher frequencies.

At the maximum intracavity power of 71 W, the 120 μW of collected THz power corresponds to a conversion efficiency of $1.7 \times 10^{-6}$ with respect to the intracavity power, and $5.6 \times 10^{-6}$ with respect to the diode pump power of only 21.4 W. We note that, owing to the standing-wave cavity geometry, THz radiation is emitted in both directions of the crystal, and only one direction is collected and characterized here; the total generated THz power is therefore twice the reported value. The acquisition time of 156 s was chosen to reach a meaningful DR within one to two minutes; obtaining 60 dB at this low level of THz average power within such a short measurement time also reflects the good stability of the driving oscillator.

## 4. Conclusion and outlook

We have demonstrated broadband intracavity THz generation using a 50-μm-thick LN plate inside a compact, low-cost diode-pumped KLM Yb:CALGO bulk oscillator. Stable 83-fs operation at 85 MHz with up to 71 W of intracavity average power was achieved using only 21.4 W of low-cost multimode diode pump power, generating single-cycle THz pulses with a bandwidth extending to 3 THz, a peak DR of 60 dB, and up to 120 μW of average power collected from one side of the emitter. The measured spectra are well reproduced by simulations of the generation and detection chain. The near-quadratic power scaling and the absence of thermal or nonlinear instabilities demonstrate that thin LN is a promising nonlinear material for intracavity operation, in contrast to the previously used GaP.

Several straightforward improvements could provide substantially higher THz power and bandwidth. Since no saturation, instability, or damage was observed up to the maximum available pump power, the intracavity power could be scaled further using higher diode pump power and a lower-loss output coupler. Such scaling would, however, require a cavity design capable of tolerating the correspondingly higher circulating peak power and nonlinear phase accumulation. In the present resonator, where the gain crystal simultaneously acts as the Kerr medium, this could be achieved by enlarging the laser mode in the crystal, at the cost of a higher lasing threshold, or by employing a thinner gain crystal with a higher doping concentration. Alternatively, a resonator with separate gain and Kerr media would allow the nonlinearity and the gain to be controlled independently [31, 32]. Ultimately, the output coupler could be replaced by a highly reflective mirror to maximize the intracavity power, while the parasitic second-harmonic radiation generated in the thin-LN plate and transmitted through the cavity mirrors could be used for laser diagnostics and as the EOS probe. Alternatively, extending the oscillator cavity to reduce the repetition rate from 85 MHz toward the 10–20 MHz range, for example using a multipass cavity extension [33], would increase the pulse energy and peak power at constant average power and thereby improve the OR conversion efficiency while retaining MHz-rate operation. The optimum repetition rate would be determined by the trade-off between conversion efficiency, cavity complexity, footprint, and the maximum tolerable intracavity peak intensity. Further gains could be obtained by optimizing the LN thickness and the beam waist on the crystal, collecting the THz radiation emitted from both sides of the plate, and purging the THz beam path with dry air or nitrogen.

Several improvements can result in a higher THz bandwidth. As shown in Fig. 3, the spectrum generated in the thin-LN plate extends well beyond the detected one: the simulated generation bandwidth is limited only by the driving pulse duration and the thin-plate phase matching, whereas the detected spectrum is additionally narrowed by the EOS response of the 3-mm GaP crystal and, at low frequencies, by the spectral filtering of the holed collection parabola. As the simulation in Fig. 3b shows, the roll-off above 2 THz is dominated by the detection crystal, so that a thinner GaP crystal is the most direct route to a broader detected bandwidth, while an alternative THz collection geometry avoiding the intracavity through-hole would mainly recover the low-frequency content and the collected power. In addition, shorter driving pulses combined with a thinner LN plate would extend the generated spectrum. The broad gain bandwidth of Yb:CALGO supports sub-30-fs Kerr-lens mode-locking under multimode diode pumping [34]. For higher THz output power slab-based Cherenkov-type LN geometries with a silicon output coupler [35] could potentially be adapted to intracavity operation. In these geometries, the THz wave exits the crystal non-collinearly with the pump, simplifying its extraction and potentially removing the need for an intracavity holed parabolic mirror.

On the detection side, the present 1-Hz delay scan was selected to provide robust lock-in detection and a high signal-to-noise ratio rather than maximum acquisition speed. Faster scanning at the present modulation frequency would therefore reduce the available integration time per delay point and could lead either to a reduced detectable THz bandwidth or to increased noise [36]. Faster acquisition would require either a higher modulation frequency or omitting the mechanical chopper and recording the balanced-detector signal continuously with a rapidly oscillating delay line and coherent averaging performed in post-processing. More fundamentally, dual-comb or ASOPS-based detection schemes [37] could eliminate the mechanical delay altogether and enable rapid, scan-free THz-TDS. The high repetition rate and passive stability of the present oscillator make it particularly well suited to such approaches in future implementations.

It is instructive to compare this approach with the fiber-coupled photoconductive systems that dominate commercial THz-TDS today. Driven by compact Er-fiber lasers, these systems are turnkey, alignment-free, and robust enough to be used outside an optics laboratory, and state-of-the-art versions now reach milliwatt-level THz average power with DR beyond 130 dB [10]. In its present free-space implementation, the source demonstrated here does not compete with such systems in cost, compactness, or ease of use, and making it transportable would require sealed and mechanically stabilized housing that is common for commercial ultrafast oscillators. Its interest lies instead in the potential scaling behavior and the low cost and simplicity of the emitter: photoconductive antennas saturate at high optical excitation, whereas OR in thin LN has been operated at hundreds of watts to kilowatts of driving average power without saturation or damage [22, 23], and the emitter itself is a mass-produced wafer that requires no lithographic fabrication. The present system is proof of principle, and adapting high-power emitter geometries such as tilted-pulse-front or Cherenkov-type configurations to intracavity operation, together with the improvements discussed above, should give access to the range of tens to hundreds of milliwatts of THz average power, which is above what photoconductive emitters currently provide. Recent mode-locked Yb:CALGO bulk oscillators delivering up to 40 W of average output power [38] indicate that the average-power handling of this gain medium is far from exhausted. A further attraction of the intracavity geometry is that the driving oscillator is itself part of the instrument: combined with a dual-comb configuration for scan-free detection, it would merge the ultrafast laser, the THz emitter, and the time-domain spectrometer into a single compact system. Such a source would be attractive for laboratory THz time-domain spectroscopy and imaging at MHz repetition rates, where the available THz power sets the achievable acquisition rate.

In summary, this simple and cost-effective architecture requires neither thin-disk technology nor active cavity stabilization. With further optimization of the intracavity power, cavity repetition rate, pulse duration, emitter geometry, THz collection, and detection scheme, it provides a practical route toward milliwatt-level, broadband, MHz-repetition-rate THz spectroscopy systems accessible to standard laser laboratories.

## Statements and Declarations

**Funding.** Deutsche Forschungsgemeinschaft (287022738 TRR 196, SFB/TRR 196, Project S02). These results are part of a project that has received funding from the European Research Council (ERC) under the European Union's HORIZON-ERC-POC programme (Project 101293326 — InTera).

**Competing Interests.** The authors declare no conflicts of interest.

**Data availability.** Data underlying the results presented in this paper are openly available in Zenodo [39].

### References

1. Koch M, Mittleman DM, Ornik J, Castro-Camus E (2023) Terahertz time-domain spectroscopy. Nat Rev Methods Primers 3:48. https://doi.org/10.1038/s43586-023-00232-z

2. Leitenstorfer A, Moskalenko AS, Kampfrath T, Kono J, Castro-Camus E, Peng K, Qureshi N, Turchinovich D, Tanaka K, Markelz AG, Havenith M, Hough C, Joyce HJ, Padilla WJ, Zhou B, Kim K-Y, Zhang X-C, Jepsen PU, Dhillon S, Vitiello M, Linfield E, Davies AG, Hoffmann MC, Lewis R, Tonouchi M, Klarskov P, Seifert TS, Gerasimenko YA, Mihailovic D, Huber R, Boland JL, Mitrofanov O, Dean P, Ellison BN, Huggard PG, Rea SP, Walker C, Leisawitz DT, Gao JR, Li C, Chen Q, Valušis G, Wallace VP, Pickwell-MacPherson E, Shang X, Hesler J, Ridler N, Renaud CC, Kallfass I, Nagatsuma T, Zeitler JA, Arnone D, Johnston MB, Cunningham J (2023) The 2023 terahertz science and technology roadmap. J Phys D: Appl Phys 56:223001. https://doi.org/10.1088/1361-6463/acbe4c

3. Buldt J, Stark H, Müller M, Grebing C, Jauregui C, Limpert J (2021) Gas-plasma-based generation of broadband terahertz radiation with 640 mW average power. Opt Lett, 46:5256–5259. https://doi.org/10.1364/OL.442374

4. Vogel T, Mansourzadeh S, Saraceno CJ (2024) Single-cycle, 643 mW average power terahertz source based on tilted pulse front in lithium niobate. Opt Lett, 49:4517. https://doi.org/10.1364/OL.532219

5. Mansourzadeh S, Vogel T, Omar A, Biggs MF, Ho ES-H, Hoberg C, Michaelis DJ, Havenith M, Johnson JA, Saraceno CJ (2025) High-dynamic-range broadband terahertz time-domain spectrometer based on organic crystal MNA. Photon Res, 13:2510–2519. https://doi.org/10.1364/PRJ.553652

6. Meyer F, Hekmat N, Vogel T, Omar A, Mansourzadeh S, Fobbe F, Hoffmann M, Wang Y, Saraceno CJ (2019) Milliwatt-class broadband THz source driven by a 112 W, sub-100 fs thin-disk laser. Opt Express, 27:30340–30349. https://doi.org/10.1364/OE.27.030340

7. Hebling J, Almási G, Kozma IZ, Kuhl J (2002) Velocity matching by pulse front tilting for large-area THz-pulse generation. Opt Express 10:1161–1166. https://doi.org/10.1364/OE.10.001161

8. Meyer F, Vogel T, Ahmed S, Saraceno CJ (2020) Single-cycle, MHz repetition rate THz source with 66 mW of average power. Opt Lett 45:2494–2497. https://doi.org/10.1364/OL.386305

9. Huiskes MD, Khalili M, Bueno J, Llombart N, Sberna PM, Saraceno CJ, Neto A (2026) High THz Average Power Generation from Photoconductive Connected Array Antenna Excited by High-Power Laser. IEEE Trans THz Sci Technol 1–11. https://doi.org/10.1109/TTHZ.2026.3669435

10. Dohms A, Vieweg N, Breuer S, Heßelmann T, Herda R, Regner N, Keyvaninia S, Gruner M, Liebermeister L, Schell M, Kohlhaas RB (2024) Fiber-Coupled THz TDS System With mW-Level THz Power and up to 137-dB Dynamic Range. IEEE Trans THz Sci Technol 14:857–864. https://doi.org/10.1109/TTHZ.2024.3467173

11. Hale LL, Jung H, Gennaro SD, Briscoe J, Harris CT, Luk TS, Addamane SJ, Reno JL, Brener I, Mitrofanov O (2022) Terahertz Pulse Generation from GaAs Metasurfaces. ACS Photonics 9:1136–1142. https://doi.org/10.1021/acsphotonics.1c01908

12. Jung H, Hale LL, Gennaro SD, Briscoe J, Iyer PP, Doiron CF, Harris CT, Luk TS, Addamane SJ, Reno JL, Brener I, Mitrofanov O (2022) Terahertz Pulse Generation with Binary Phase Control in Nonlinear InAs Metasurface. Nano Lett 22:9077–9083. https://doi.org/10.1021/acs.nanolett.2c03456

13. Pupeza I, Zhang C, Högner M, Ye J (2021) Extreme-ultraviolet frequency combs for precision metrology and attosecond science. Nat Photonics 15:175–186. https://doi.org/10.1038/s41566-020-00741-3

14. Fischer J, Drs J, Labaye F, Modsching N, Müller M, Wittwer VJ, Südmeyer T (2022) Efficient XUV-light out-coupling of intra-cavity high harmonics by a coated grazing-incidence plate. Opt Express 30:30969–30979. https://doi.org/10.1364/OE.458946

15. Darmo J, Müller T, Strasser G, Unterrainer K, Le T, Stingl A, Tempea G (2002) Voltage-controlled intracavity terahertz generator for self-starting Ti:sapphire lasers. Opt Lett 27:1941. https://doi.org/10.1364/OL.27.001941

16. Matthäus G, Ortaç B, Limpert J, Nolte S, Hohmuth R, Voitsch M, Richter W, Pradarutti B, Tünnermann A (2008) Intracavity terahertz generation inside a high-energy ultrafast soliton fiber laser. Appl Phys Lett 93:261105. https://doi.org/10.1063/1.3056118

17. Theuer M, Molter D, Maki K, Otani C, L'huillier JA, Beigang R (2008) Terahertz generation in an actively controlled femtosecond enhancement cavity. Appl Phys Lett 93:041119. https://doi.org/10.1063/1.2966342

18. Xu S, Liu J, Zheng G, Li J (2010) Broadband terahertz generation through intracavity nonlinear optical rectification. Opt Express 18:22625. https://doi.org/10.1364/OE.18.022625

19. Hamrouni M, Drs J, Modsching N, Wittwer VJ, Labaye F, Südmeyer T (2021) Intra-oscillator broadband THz generation in a compact ultrafast diode-pumped solid-state laser. Opt Express 29:23729. https://doi.org/10.1364/OE.426750

20. Hekmat N, Vogel T, Wang Y, Mansourzadeh S, Aslani F, Omar A, Hoffmann M, Meyer F, Saraceno CJ (2020) Cryogenically cooled GaP for optical rectification at high excitation average powers. Opt Mater Express 10:2768–2782. https://doi.org/10.1364/OME.402564

21. Fülöp JA, Tzortzakis S, Kampfrath T (2020) Laser-Driven Strong-Field Terahertz Sources. Adv Opt Mater 8:1900681. https://doi.org/10.1002/adom.201900681

22. Wang Y, Vogel T, Khalili M, Mansourzadeh S, Hasse K, Suntsov S, Kip D, Saraceno CJ (2023) High-power intracavity single-cycle THz pulse generation using thin lithium niobate. Optica 10:1719–1722. https://doi.org/https://doi.org/10.1364/OPTICA.504513

23. Suerra E, Canella F, Giannotti D, Khalili M, Wang Y, Hasse K, Suntsov S, Kip D, Saraceno C, Cialdi S, Galzerano G (2025) Ytterbium-laser-driven THz generation in thin lithium niobate at 1.9 kW average power in a passive enhancement cavity. APL Photonics 10:046111. https://doi.org/10.1063/5.0252040

24. Yefet S, Pe'er A (2013) A Review of Cavity Design for Kerr Lens Mode-Locked Solid-State Lasers. Applied Sciences 3:694–724. https://doi.org/10.3390/app3040694

25. van Exter M, Fattinger Ch, Grischkowsky D (1989) Terahertz time-domain spectroscopy of water vapor. Opt Lett 14:1128. https://doi.org/10.1364/OL.14.001128

26. Hattori T, Takeuchi K (2007) Simulation study on cascaded terahertz pulse generation in electro-optic crystals. Opt Express 15:8076. https://doi.org/10.1364/OE.15.008076

27. Unferdorben M, Szaller Z, Hajdara I, Hebling J, Pálfalvi L (2015) Measurement of Refractive Index and Absorption Coefficient of Congruent and Stoichiometric Lithium Niobate in the Terahertz Range. J Infrared Millim Terahertz Waves 36:1203–1209. https://doi.org/10.1007/s10762-015-0165-5

28. Faure J, Van Tilborg J, Kaindl RA, Leemans WP (2004) Modelling Laser-Based Table-Top THz Sources: Optical Rectification, Propagation and Electro-Optic Sampling. Opt Quantum Electron 36:681–697. https://doi.org/10.1023/B:OQEL.0000039617.85129.c2

29. Wu X, Zhou C, Huang WR, Ahr F, Kärtner FX (2015) Temperature dependent refractive index and absorption coefficient of congruent lithium niobate crystals in the terahertz range. Opt Express 23:29729. https://doi.org/10.1364/OE.23.029729

30. Barker AS, Loudon R (1967) Dielectric Properties and Optical Phonons in LiNbO3. Phys Rev 158:433–445. https://doi.org/10.1103/PhysRev.158.433

31. Ishikawa T, Eilanlou AA, Nabekawa Y, Fujihira Y, Imahoko T, Sumiyoshi T, Kannari F, Kuwata-Gonokami M, Midorikawa K (2015) Kerr lens mode-locked Yb:$Lu_2O_3$ bulk ceramic oscillator pumped by a multimode laser diode. Jpn J Appl Phys 54:072703. https://doi.org/10.7567/JJAP.54.072703

32. Meyer JG, Khalili M, Pronin O (2023) Systematic study of peak power scaling for an Yb-doped Kerr-lens mode-locked bulk oscillator. Appl Phys B 129:161. https://doi.org/10.1007/s00340-023-08104-w

33. Saraceno CJ, Emaury F, Schriber C, Hoffmann M, Golling M, Südmeyer T, Keller U (2014) Ultrafast thin-disk laser with 80 µJ pulse energy and 242 W of average power. Opt Lett 39:9. https://doi.org/10.1364/OL.39.000009

34. Trawi F, Drs J, Müller M, Hamrouni M, Wittwer VJ, Südmeyer T (2024) Sub-30-fs Yb:CALGO laser oscillator based on cross-polarized multi-mode diode pumping. Optics Express 32:37897–37905. https://doi.org/10.1364/OE.537753

35. Tsarev MV, Ehberger D, Baum P (2016) High-average-power, intense THz pulses from a LiNbO3 slab with silicon output coupler. Appl Phys B 122:30. https://doi.org/10.1007/s00340-015-6315-6

36. Vogel T, Saraceno CJ (2024) Advanced Data Processing of THz-Time Domain Spectroscopy Data with Sinusoidally Moving Delay Lines. J Infrared Millim Terahertz Waves 45:967–983. https://doi.org/https://doi.org/10.1007/s10762-024-01012-w

37. Willenberg B, Phillips CR, Pupeikis J, Camenzind SL, Liebermeister L, Kohlhass RB, Globisch B, Keller U (2024) THz-TDS with gigahertz Yb-based dual-comb lasers: noise analysis and mitigation strategies. Appl Opt 63:4144–4156. https://doi.org/10.1364/AO.522802

38. Su X, Wang Y, Qu Y, Zhang B, Niu B, Zhou P, He J, Zhang B (2026) 40-W high-power femtosecond mode-locking Yb:CALGO bulk oscillator. Photon Res 14:4026. https://doi.org/10.1364/PRJ.605991

39. Khalili M, Pimpė J, Wang Y, Vengelis J, Hasse K, Suntsov S, Kip D, Saraceno CJ (2026) Intracavity THz generation using a thin lithium niobate plate in a compact Kerr-lens mode-locked Yb:CALGO bulk oscillator. Zenodo. https://doi.org/10.5281/zenodo.21650566